# Hypergraph modelling of wave scattering to speed-up material design

Kunwoo Park[1], Ikbeom Lee[1], Seungmok Youn[1], Gitae Lee[1], Namkyoo Park[2,§], and Sunkyu Yu[1,*]

[1]Intelligent Wave Systems Laboratory, Department of Electrical and Computer Engineering, Seoul National University, Seoul 08826, Korea

[2]Photonic Systems Laboratory, Department of Electrical and Computer Engineering, Seoul National University, Seoul 08826, Korea

E-mail address for correspondence: [§]nkpark@snu.ac.kr; [*]sunkyu.yu@snu.ac.kr

**Abstract**

Hypergraphs offer a generalized framework for understanding complex systems, covering group interactions of different orders beyond traditional pairwise interactions. This modelling allows for the simplified description of simultaneous interactions among multiple elements in coupled oscillators, graph neural networks, and entangled qubits. Here, we employ this generalized framework to describe wave-matter interactions for material design acceleration. By devising the set operations for multiparticle systems, we develop the hypergraph model, which compactly describes wave interferences among multiparticles in scattering events by hyperedges of different orders. This compactness enables an evolutionary algorithm with $O(N^{1/2})$ time complexity and approximated accuracy for designing stealthy hyperuniform materials, which is superior to traditional methods of $O(N)$ scaling. By hybridizing our hypergraph evolutions to the conventional collective-coordinate method, we preserve the original accuracy, while achieving substantial speed-up in approaching near the optimum. Our result paves the way toward scalable material design and compact interpretations of large-scale multiparticle systems.

# Introduction

Employing a statistical perspective on material science has substantially extended material phases characterized by microstructures[1]. Beyond the traditional classification of ordered and disordered phases that are represented by crystals and Poisson materials, respectively, the concept of correlated disorder has unveiled various counterintuitive wave phenomena[2]. For example, utilizing the reciprocal-space design of two-point correlation functions, hyperuniformity[3] and stealthiness[4] enable the coexistence of wave behaviours traditionally considered those in crystals and in Poisson materials: perfect bandgaps and transparency with statistically isotropic responses[5,6].

To exploit such statistically defined material phases, it is crucial to develop efficient design strategy of large-scale systems for satisfying preconditions, such as statistical homogeneity and ergodicity[1]. Representative approaches include variational methods[7-9], evolutionary designs[10-12], and deep learning[13,14]. In all these methods, the critical issue lies in realizing the fast and accurate design process. Although the methods based on governing equations offer relatively high accuracy, their time complexity usually has a lower bound of $O(N)$, where $N$ denotes the number of elements. Recently, deep learning has attracted much attention because of excellent speed-up at a given $N$, despite its relatively degraded accuracy[13]. However, the increase of $N$ in deep learning based on matrix operations requires severely degraded scaling of $\sim O(N^2)$ in preparing datasets, training neural networks, and conducting inferences.

To resolve this issue, we can envisage reducing the effective number of elements for a given $N$. This reduction can be accomplished by grouping related sub-elements across the entire system—an approach analogous to hypergraphs in network science[15-17]. The concept of hypergraphs has enabled the representation of interactions among groups of elements—

hyperedges—which provides the compact descriptions of long-range entangled qubits[18,19], neural network weights[20,21], and social interactions[22,23]. By introducing hypergraphs to material design, we can expect the representation of collective wave-matter interactions of a group of elements by a hyperedge, thereby allowing for the reduction of design parameters.

Here, we develop the hypergraph modelling of wave scattering, which we call scattering hypergraph, to accelerate material design. For multiparticle systems, we define set operations to generate a large-scale system from seed subsets. Using these set operations under the Born approximation, we model interferences among each particle group as hyperedges of different orders. The obtained hypergraph represents scattering events with greatly reduced design parameters, allowing for $O(N^{1/2})$ time complexity in our example of designing stealthy hyperuniform (SHU) materials. By combining this model to the collective-coordinate method[7], we simultaneously achieve the substantial speed-up of material design while preserving its accuracy. Our hypergraph model with the hybrid design method offers not only remarkable performance figures in material design but also a viewpoint on capturing collective wave-matter interactions.

## Results

**Set operations for scattering hypergraphs**

Consider an $N$-particle system composed of non-interacting identical point particles. The system is characterized by a set of particle positions, as $\mathbf{R} = \{\mathbf{r}_n: 1 \leq n \leq N\}$, where $\mathbf{r}_n$ is the position of the $n$th particle. For a planewave incidence under the first-order Born approximation, the scattering intensity is proportional to the structure factor[24], $S(\mathbf{k}) = |\Sigma_n \exp(-i\mathbf{k}\cdot\mathbf{r}_n)|^2/N$, where $\mathbf{k} = \mathbf{k}_s - \mathbf{k}_i$ is the wavevector shift between the incident and scattered wavevectors $\mathbf{k}_i$ and $\mathbf{k}_s$, respectively (Fig. 1a).

As demonstrated in the previous work[12], the overall scattering response of the system can be modelled with a graph network, which possesses self- and pairwise interactions, or equivalently, first- and second-order interactions. In this modelling, a node and an edge of the graph correspond to a particle and the scattering interference between a pair of particles, respectively, leading to the following edge weight between the $n$th and $n'$th nodes:

$$w_{n,n'} = \left\langle \cos\left[\mathbf{k} \cdot (\mathbf{r}_n - \mathbf{r}_{n'})\right] \right\rangle_{\mathbf{K}} \triangleq w_m, \qquad (1)$$

where $m$ denotes the index of the edge ($n \neq n'$, $1 \leq m \leq M$ for $M = N(N-1)/2$), $\mathbf{K}$ is the region of interest in reciprocal space, and $\langle \ldots \rangle_{\mathbf{K}}$ denotes the average inside $\mathbf{K}$ (Fig. 1b). According to Eq. (1), this weighted graph allows for describing wave scattering using network parameters, as follows:

$$\langle S(\mathbf{k}) \rangle_{\mathbf{K}} = 1 + \frac{2}{N}\sum_{m=1}^{M} w_m = 1 + \frac{1}{N}\sum_{n=1}^{N} d_n, \qquad (2)$$

where $d_n$ is the node degree of the $n$th node. As examined in network science[25], the node degree represents the centrality of each node, quantifying the impact of the node on the entire scattering. Despite the advantages of this graph modelling, such as the evolutionary design[12] analogous to scale-free networks[26], its computational efficiency is equivalent to traditional scattering theory: $O(N)$ scaling in evaluating $N$ node degrees.

Revisiting critical advantages of hypergraphs in describing collective dynamics in social[22,23], biological[27,28], chemical[29], and quantum systems[18,19], we employ the concept of hypergraphs to the modelling of scattering events, focusing on grouping shared interactions, and therefore, achieving below $O(N)$ computation efficiency. In this context, inspired by the hyperedge in hypergraph theory[15], which is classified by its order—the number of connected nodes—we introduce the concept of the hyperedge classified by the number of participating particles in the scattering event.

To characterize participating particles in a hyperedge, we introduce a seed set of particle positions $\mathbf{Q} = \{\mathbf{r}_q: 1 \leq q \leq N_\mathbf{Q}\}$ (Fig. 1c) and its subsets $\mathbf{Q}_l \subset \mathbf{Q}$, each composed of $N_l$ particles ($1 \leq l \leq L$; Fig. 1d), where $\mathbf{Q}_l$ are mutually exclusive sets satisfying $\mathbf{Q} = \cup_l \mathbf{Q}_l$ and $N_\mathbf{Q} = \Sigma_l N_l$. As the mathematical foundation of defining hyperedges, we develop three fundamental set operations: union ($\mathbf{Q}_1 \cup \mathbf{Q}_2$, Fig. 1e), complement ($\mathbf{Q}_1 \setminus \mathbf{Q}_2$, Fig. 1f), and Minkowski addition ($\mathbf{Q}_1 + \mathbf{Q}_2$, Fig. 1g). By combining these set operations, we can obtain diverse particle systems distinct from $\mathbf{Q}$, possessing different particle distributions and unpreserved particle number. Figure 1h shows an example of resulting systems from $\mathbf{R} = (\mathbf{Q}_1 + \mathbf{Q}_2) \cup \mathbf{Q}_3$, which includes $N = N_1 N_2 + N_3$ particles. When considering the particle number of $\mathbf{Q}$, $N_\mathbf{Q} = N_1 + N_2 + N_3$, we can expect the systematic design of large-scale systems from a small seed system.

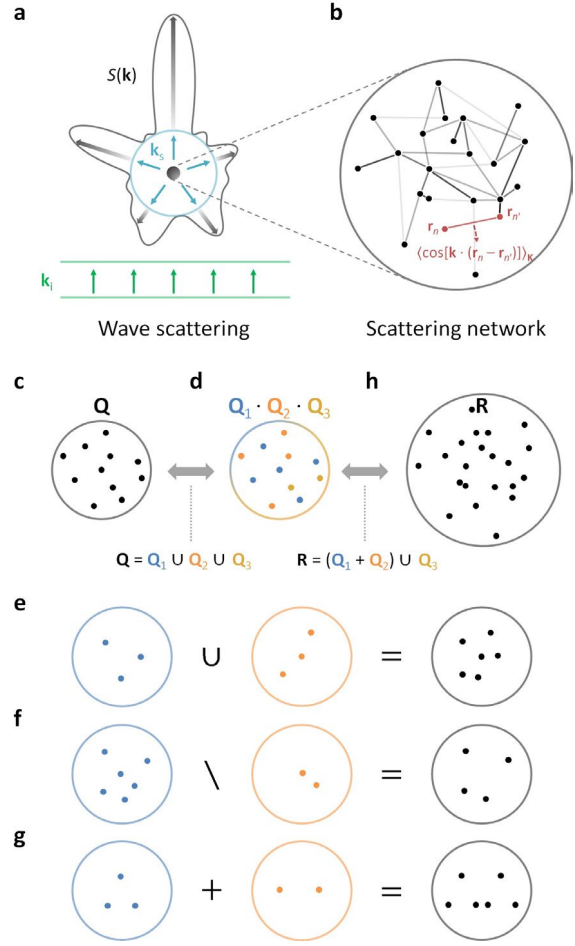

**Fig. 1. Set operations for scattering hypergraphs. a**, Scattering from a multiparticle system. The black circle represents the multiparticle system. Green and cyan arrows represent incident and scattered waves with wavevector $\mathbf{k}_i$ and $\mathbf{k}_s$, respectively. The grey line indicates the structure factor of the system, while the lengths of the grey arrows describe the angular scattering intensity. **b**, The scattering network modelling of the system. Each dot and solid line represent a node and an edge, respectively. Only a small portion of the edges are shown. The transparency of the solid lines indicates the edge weights. The red dots represent the $n$th and $n'$th nodes, and the red solid line represents the edge connecting the two nodes. **c,d,h**, The seed set $\mathbf{Q}$ (**c**) decomposed into three subsets $\mathbf{Q}_1$, $\mathbf{Q}_2$, and $\mathbf{Q}_3$ (**d**) and the material phase $\mathbf{R}$ satisfying $\mathbf{R} = (\mathbf{Q}_1 + \mathbf{Q}_2) \cup \mathbf{Q}_3$ (**h**). **e-g**, Three types of set operations: the union (**e**), complement (**f**), and Minkowski addition (**g**).

## Scattering hypergraphs

Under the set operations described in Fig. 1, we develop a hypergraph model to describe scattering from a multiparticle system $\mathbf{R}$, where the nodes of the hypergraph correspond to particles of the seed system $\mathbf{Q} = \cup_l \mathbf{Q}_l$, and the hyperedges are characterized by the participating set operations. Notably, there are an infinite number of allowed combinations of the set operations as described in Fig. 2a, each of which represents a distinct multiparticle system. As an example, we illustrate the hypergraph modelling for the relation $\mathbf{R} = (\mathbf{Q}_1 + \mathbf{Q}_2) \cup \mathbf{Q}_3$ (Fig. 2b,c). By examining the scattering from the system $\mathbf{R}$ under the Born approximation (Fig. 2b), the scattering intensity, which is identical to the structure factor $S(\mathbf{k})$, can be expressed as (Supplementary Note S1):

$$\langle S(\mathbf{k}) \rangle_{\mathbf{K}} = 1 + \frac{2}{N} \sum_i \eta_i \sum_{m=1}^{M_i} w_m^i, \tag{3}$$

where $i$ and $m$ denote the type and index of hyperedges, respectively, $M_i$ is the number of the $i$-type hyperedges, $w_m^i$ denotes the weight of the $m$th $i$-type hyperedge, and $\eta_i$ represents the group weight of $i$-type hyperedges.

Equation (3) illustrates the hypergraph representation of the scattering from the material $\mathbf{R}$. In the hypergraph, hyperedges are classified by their type $i = (l_1, l_2, \ldots)$, which depicts the subsystems of participating nodes in each hyperedge. While the allowed types are determined by the governing set operation, here, $\mathbf{R} = (\mathbf{Q}_1 + \mathbf{Q}_2) \cup \mathbf{Q}_3$ (Fig. 2c; Supplementary Note S1), each type possesses its unique expression of a hyperedge weight (Supplementary Tables S1-S3). For example, the type of $i = (1, 2, 3)$ dictates the participating nodes of $\mathbf{r}_n \in \mathbf{Q}_1$, $\mathbf{r}_{n'} \in \mathbf{Q}_2$, and $\mathbf{r}_{n''} \in \mathbf{Q}_3$ (Fig. 2c). According to the set operation $\mathbf{R} = (\mathbf{Q}_1 + \mathbf{Q}_2) \cup \mathbf{Q}_3$, the resulting particles in $\mathbf{R}$ are located at $\mathbf{r}_n + \mathbf{r}_{n'}$ and $\mathbf{r}_{n''}$. The scattering interference from these particles (Fig. 2b) leads to the following third-order hyperedge (Fig. 2c):

$$w_m^{(1,2,3)} = \langle \cos[\mathbf{k} \cdot (\mathbf{r}_n + \mathbf{r}_{n'} - \mathbf{r}_{n''})] \rangle_{\mathbf{K}}, \tag{4}$$

where $m$ is the index of the (1,2,3)-type hyperedge ($1 \leq m \leq M_{(1,2,3)} = N_1 N_2 N_3$) for the set ($\mathbf{r}_n$, $\mathbf{r}_{n'}$, $\mathbf{r}_{n''}$). Because both pairwise- and hyper-graph models represent the same scattering phenomenon, we can always find the corresponding pairwise edges for each hyperedge, while demonstrating the superior compactness of the hypergraph model (Supplementary Note S2).

As the generalization of node degrees in network science[25], we define the $i$-type hyperdegree $d_n^i$ by summing all the $i$-type hyperedge weight at each node, which quantifies the impact of the $n$th node in **Q** on the $i$-type scattering (Methods). While a node in **Q** imposes differentiated impacts on each type of hyperedges, the contribution of each hyperedge type to the overall scattering is also differentiated, which is characterized by the group weight $\eta_i$. Such an increased expressivity of network structures allows for the representation of diverse multiparticle systems **R** using a small seed system **Q**.

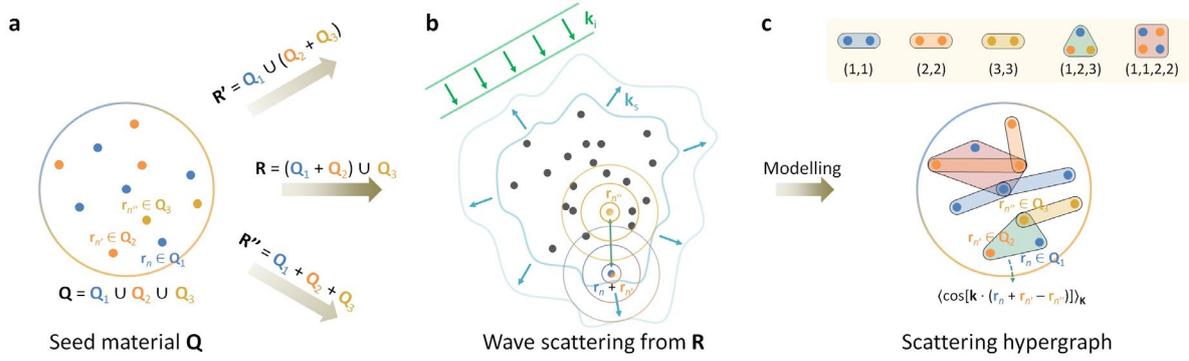

**Fig. 2. Scattering hypergraph configuration. a**, A seed material $\mathbf{Q} = \mathbf{Q}_1 \cup \mathbf{Q}_2 \cup \mathbf{Q}_3$, allowing for generating various material phases through set operations, such as **R**, **R'**, and **R"**. **b**, Wave scattering from $\mathbf{R} = (\mathbf{Q}_1 + \mathbf{Q}_2) \cup \mathbf{Q}_3$. Green and cyan arrows represent incident and scattered waves with wavevectors $\mathbf{k}_i$ and $\mathbf{k}_s$, respectively. **c**, The scattering hypergraph for **R**. Because every group of nodes of type $i$ = (1,1), (2,2), (3,3), (1,2,3), or (1,1,2,2) is linked by an $i$-type hyperedge, only a portion of hyperedges is visualized. For example, the (1,2,3)-type hyperedge connecting three nodes at $\mathbf{r}_n \in \mathbf{Q}_1$, $\mathbf{r}_{n'} \in \mathbf{Q}_2$, and $\mathbf{r}_{n''} \in \mathbf{Q}_3$ is illustrated. Its weight is defined as Eq. (4).

**Hypergraph generation of materials**

Based on the hypergraph modelling of scattering, we demonstrate the designer generation of large-scale multiparticle systems **R** from a small-scale seed system **Q**. As an illustrative example, we investigate the set operation $\mathbf{R} = \mathbf{Q}_1 + \mathbf{Q}_2$, which enables the generation of $10^4$- particle systems with the target scattering responses by manipulating 200 particles of $N_1 = N_2 = 100$.

For the multiparticle system generation, we develop the evolutionary design process of a scattering hypergraph (Fig. 3a-d), which is constituted with the 'Growth' and 'Projection' processes. In the 'Growth' stage at each epoch, a new node is attached to one of the seed subsystems $\mathbf{Q}_1$ or $\mathbf{Q}_2$ at the optimal position to minimize a specific cost function for the target scattering (from Fig. 3a to Fig. 3c). In the 'Projection' stage, the new node is projected to a real space according to the set operation in Fig. 1, leading to the generation of multiparticles in a real space (from Fig. 3b to Fig. 3d). This emergence of multiparticles enables large-scale system generation in our hypergraph modelling.

As an example, we apply our design process to generate SHU materials[3], which are characterized by suppressed scattering responses against long-wavelength incident waves: $\langle S(\mathbf{k})\rangle_\mathbf{K} \approx 0$, where the region of interest $\mathbf{K}$ denotes a finite reciprocal space around $\mathbf{k} = 0$ (Methods). To minimize $\langle S(\mathbf{k})\rangle_\mathbf{K}$ in Eq. (3), the cost function is defined as (Supplementary Note S3):

$$\rho(\mathbf{r};l) = \sum_i \eta_i d^i_{\text{evol}}(\mathbf{r};l), \qquad (5)$$

where $d^i_{\text{evol}}(\mathbf{r};l)$ denotes the $i$-type hyperdegree of a node located at $\mathbf{r}$ in the $l$th subsystem $\mathbf{Q}_l$ ($l = 1$ or 2). When attaching new nodes and evaluating $S(\mathbf{k})$, we employ the periodic boundary condition with a supercell configuration (Methods).

Figures 3e and 3f illustrate the evolutions of hypergraph nodes $\mathbf{Q} = \mathbf{Q}_1 \cup \mathbf{Q}_2$ and the corresponding multiparticle system $\mathbf{R}$, respectively, and the resulting evolution $\langle S(\mathbf{k})\rangle_\mathbf{K}$ is shown in Fig. 3g (Supplementary Note S4 for other set operations). The evolutions show that the optimization process operates well with rapid reduction of $\langle S(\mathbf{k})\rangle_\mathbf{K}$ before experiencing the finite boundary of the supercell near the epoch $t = 22$. Although the overall trend of decreasing $\langle S(\mathbf{k})\rangle_K$ maintains also after $t = 22$, the speed of $\langle S(\mathbf{k})\rangle_K$ suppression becomes substantially reduced, hindering the ultimate approach to the desired level of suppression (Supplementary Video S1 for the time evolutions of material systems **Q** and **R**).

To investigate the origin of this phenomenon, we examine the effect of each hyperedge, which can be quantified by $S_i = \Sigma_m 2\eta_i w_m^i/N$ ($i = (1,1)$, $(2,2)$, or $(1,1,2,2)$), where $\langle S(\mathbf{k})\rangle_K = 1 + \Sigma_i S_i$ from Eq. (3). Figure 3h shows the evolution of each edge component, representing their differentiated contributions to the total cost function. We note that such contributions are tunable by imposing the preference on each hyperedge (Supplementary Note S5). However, the existence of undesired contributions, here from $S_{(1,1,2,2)}$, hinders the complete suppression of wave scattering in **R**, and therefore, the realization of the target material. Considering the handling of large-scale systems from the 'Projection' process and the rapid decrease before experiencing boundary effects, we can characterize the properties of hypergraph material generation—speed-up but rough design process to the optimum.

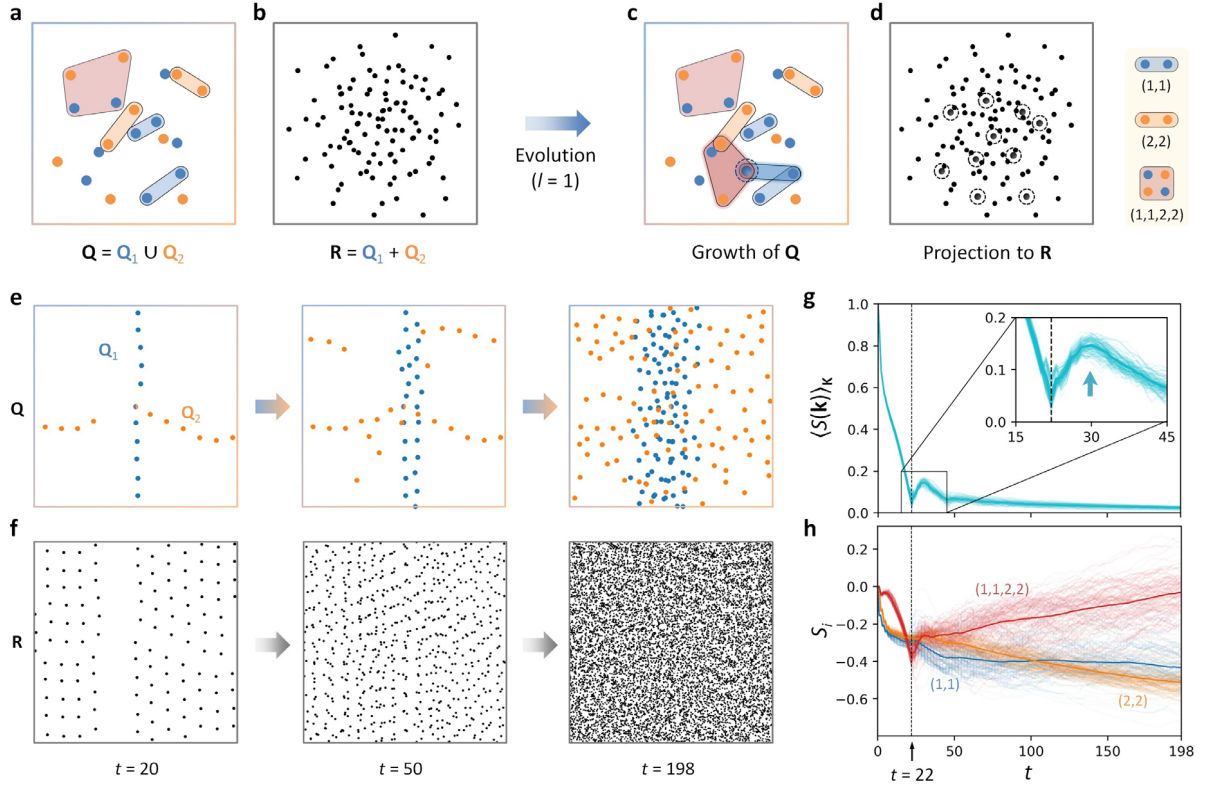

**Fig. 3. Hypergraph design of SHU materials. a,c**, Scattering hypergraphs before (**a**) and after (**c**) the attachment of a new node of $Q_1$ from the 'Growth' process. The spatial configuration of the nodes represents the seed system $Q = Q_1 \cup Q_2$. **b,d**, Generated materials $R = Q_1 + Q_2$ before (**b**) and after (**d**) the 'Projection' process. **e,f**, Evolution of material structures of $Q$ (**e**) and $R$ (**f**) at $t = 20$, 50, and 198. Because each subsystem evolves in alternating order, the particle number of $R$ is $N = (t/2 + 1)^2$ for even epoch $t$. The boxes represent the supercell. **g,h**, Time evolutions of $\langle S(\mathbf{k})\rangle_K$ (**g**) and $S_i$ (**h**). 100 realizations are investigated with different Monte Carlo sampling of a real space (Methods).

**Time complexity and hybrid design**

To estimate and advance the proposed design method, we analyse the scalability in time complexity (Fig. 4a) and accuracy (Fig. 4b) of our method compared to the traditional collective-coordinate method[7] and pairwise-edge evolving networks[12]. Figure 4a demonstrates $O(N^{1/2})$ time complexity of the proposed method in generating $N$-particle SHU systems, which is superior to $O(N)$ scaling of both the collective-coordinate method and evolving scattering networks. We note that, despite recent advances in the collective-coordinate method to accelerate large-scale SHU design[30,31], the complexity remains $O(N)$ for a fixed suppression region **K**. The observed improvement in our approach originates from the quadratic scaling of the particle number $N = O(t^2)$ for the employed set operation $\mathbf{R} = \mathbf{Q}_1 + \mathbf{Q}_2$ (Supplementary Note S6 and Supplementary Algorithm S1 for details). Therefore, we can achieve further improvements by adopting different set operations, for example, $O(N^{1/3})$ for $\mathbf{R} = \mathbf{Q}_1 + \mathbf{Q}_2 + \mathbf{Q}_3$.

However, as expected from Figs. 3g and 3h, superior scalability appears with the degraded design accuracy, as shown in $\langle S(\mathbf{k}) \rangle_\mathbf{K}$ in Fig. 4b. While evolutionary methods outperform the collective-coordinate method with uniformly random initializations for small $N$, they cannot completely suppress wave scattering for large-$N$ systems in contrast to the collective-coordinate method. Notably, hypergraph design exhibits even worse accuracy than pairwise-edge networks due to the reduction of design freedom from the hyperedge grouping of interferences.

To reconcile design speed-up and strict SHU constraints, we develop a hybrid method that combines scattering hypergraphs and collective-coordinate methods, analogous to quantum-classical hybrid acceleration in reinforcement learning[32]. At the first stage of the design, scattering hypergraph evolves to achieve an $N$-particle quasi-SHU system with $O(N^{1/2})$ time scaling. In the subsequent stage, we apply collective-coordinate optimization to fine-tune the obtained system for the strict SHU condition. Figure 4c shows superior design performance of the proposed hybrid design compared to the stand-alone collective-coordinate method.

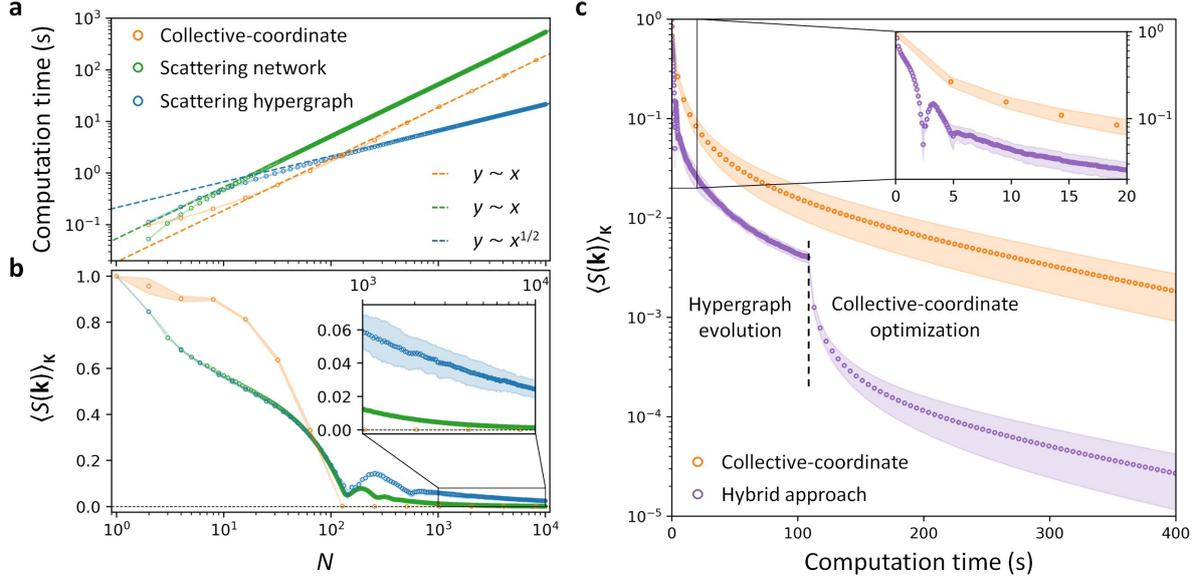

**Fig. 4. Time complexity analysis. a,b,** Computation time for the design of an *N*-particle material (**a**) and resulting scattering response $\langle S(\mathbf{k}) \rangle_\mathbf{K}$ (**b**) using the collective-coordinate method (orange) and time evolution of scattering networks (green) and scattering hypergraphs (blue). **c,** Temporal variation of $\langle S(\mathbf{k}) \rangle_\mathbf{K}$ during the optimization process from the collective-coordinate method (orange) and the hybrid approach (purple). In the hybrid design approach, hypergraph evolution is terminated at the dotted vertical line and collective-coordinate optimization follows. The final particle number is set to $N = 500^2$ in both methods for comparison. Circular points and error bands denote the ensemble average and the standard deviation of an ensemble of 50 realizations for each design method in (**a-c**). See Methods for details.

## Discussion

When considering the rapid and highly accurate convergence of our hybrid method to the optimum, the hypergraph evolution stage in the method can be interpreted as the preparation of well-operating initial conditions applied to the collective-coordinate method. We note that this preparation is efficient to search for the global optimum, because of the evolutionary nature of the proposed hypergraph design starting from one particle that possesses the minimum initial constraint. This idea indicates a general recipe for developing a hybrid of evolving and static design procedures not restricted to our hypergraph modelling.

Although we have demonstrated the emergence of fourth-order hyperedges and design acceleration with the simplest Minkowski addition, $\mathbf{R} = \mathbf{Q}_1 + \mathbf{Q}_2$, the systematic investigation on extended set operations will fully realize the potential of our method. For instance, generalizing $\mathbf{R} = \mathbf{Q}_1 + \mathbf{Q}_2$ into $\mathbf{R} = \mathbf{Q}_1 + \mathbf{Q}_2 + \ldots + \mathbf{Q}_L$ could facilitate very-large-scale material design with $O(N^{1/L})$ time complexity through up to $2L$-order hyperedge modelling. Considering the major impact of evolving algorithms—varying system sizes and updating rules[25,33,34]—we can also envisage the dynamical change of set operations along the evolution.

In summary, we have developed a hypergraph framework for scalable design of materials with the target scattering. Based on the set operations for multiple particles and their combinations, we interpreted scattering responses through hyperedges of different orders. This hypergraph modelling substantially reduces the effective particle number representing large-scale systems, allowing for the quadratic reduction in computation time in our example. We note that such a group interaction modelling resembles the tensor-product representation in quantum states[35], which inspires the hybrid algorithm combined with conventional high-accuracy design method analogous to quantum-classical hybrid learning[32]. We envisage the extension of our perspective beyond scattering events to various wave phenomena.

## Methods

**Node hyperdegrees.** Analogous to Eq. (2) that decomposes the scattering response of the entire system into the contributions of each node—the node degree—we examine the scattering response of the material $\mathbf{R}$ through node hyperdegrees. Equation (2) follows directly from the handshaking lemma for pairwise graphs:

$$\sum_{n=1}^{N} d_n = 2 \sum_{m=1}^{M} w_m. \qquad (6)$$

The handshaking lemma is generalized for *k*-uniform hypergraphs, which only contain hyperedges of order *k*. Because scattering hypergraphs include hyperedges of diverse orders, we apply the generalized handshaking lemma to each interaction type *i*, as follows:

$$\sum_{n=1}^{N} d_n^i = |i| \sum_{m=1}^{M_i} w_m^i, \tag{7}$$

where $|i|$ denotes the order of *i*-type hyperedges. Substituting Eq. (7) into Eq. (3) yields a direct link between the total scattering response and node hyperdegrees:

$$\langle S(\mathbf{k}) \rangle_{\mathbf{K}} = 1 + \frac{2}{N} \sum_i \frac{\eta_i}{|i|} \sum_{n=1}^{N} d_n^i. \tag{8}$$

**Preconditions for SHU material design.** We impose a periodic boundary condition on **Q** and **R** with the fundamental cell $\mathbf{\Omega} = \{(x, y): -0.5 \leq x, y < 0.5\}$. We note that the Minkowski addition between two periodic subsystems $\mathbf{Q}_1$ and $\mathbf{Q}_2$ is defined the modulo periodicity, as $\mathbf{Q}_1 + \mathbf{Q}_2 = \{(x + x' - \lfloor x + x' + 0.5 \rfloor, y + y' - \lfloor y + y' + 0.5 \rfloor): (x, y) \in \mathbf{Q}_1, (x', y') \in \mathbf{Q}_2\}$, where $\lfloor \ldots \rfloor$ denotes the floor function. In designing SHU materials with suppressed long-wavelength scattering, we focus on the reciprocal space region $\mathbf{K} = \{\mathbf{k} \in \Lambda: 0 < |\mathbf{k}| \leq k_{\text{th}}\}$, where $\Lambda = (2\pi\mathbb{Z})^2$ is the reciprocal lattice. We set the threshold wavenumber $k_{\text{th}} = 20\pi$.

**Evolutionary design process.** We initialize the target material as $\mathbf{R} = \{\mathbf{0}\}$ and set both seed subsystems as $\mathbf{Q}_1 = \mathbf{Q}_2 = \{\mathbf{0}\}$. These initial conditions do not lose generality because of the invariance of the structure factor against any translational shifts of the material **R** and the application of the periodic boundary condition. During the hypergraph evolution, each subsystem $\mathbf{Q}_l$ evolves in alternating order, starting from $\mathbf{Q}_1$. The cost function is evaluated for candidate points $\mathbf{\Omega}_{\text{cand}}$ randomly sampled in the fundamental cell $\mathbf{\Omega}$ through the Monte Carlo method. The number of Monte Carlo sampling points are set to $10^4$. Because the cost function can be evaluated in $O(1)$ computation time independent of the current material size $N$ (Supplementary Note S6), the overall time complexity for hypergraph material design is $O(t) = O(N^{1/2})$ for the time step $t$. On the other hand, the pairwise graph model corresponds to a trivial set operation $\mathbf{R} = \mathbf{Q}$, exhibiting $O(t) = O(N)$ scaling. See Supplementary Algorithm S1 for the pseudo-code form of our hypergraph evolution.

**Collective-coordinate design process.** The collective-coordinate algorithm minimizes $\langle S(\mathbf{k})\rangle_\mathbf{K}$ via gradient descent. Starting from an initial random configuration $\mathbf{R} = \{\mathbf{r}_n: 1 \leq n \leq N\}$ in $\Omega$, at each epoch, the collective-coordinate variables $\mu(\mathbf{k}) = \Sigma_n \exp(-i\mathbf{k}\cdot\mathbf{r}_n)$ for all $\mathbf{k} \in \mathbf{K}$ are computed. The gradient of $\langle S(\mathbf{k})\rangle_\mathbf{K}$ for the $n$th particle is evaluated as:

$$\nabla_{\mathbf{r}_n} \langle S(\mathbf{k})\rangle_\mathbf{K} = \frac{2}{N}\left\langle \mathbf{k}\cdot \mathrm{Im}\left(e^{-i\mathbf{k}\cdot\mathbf{r}_n}\mu^*(\mathbf{k})\right)\right\rangle_\mathbf{K}. \tag{9}$$

Using the gradients for every particle, each particle position is updated by:

$$\mathbf{r}_n \to \mathbf{r}_n - \varepsilon\cdot\nabla_{\mathbf{r}_n}\langle S(\mathbf{k})\rangle_\mathbf{K}, \tag{10}$$

where $\varepsilon = 0.05$ is the learning rate. This algorithm yields $O(N)$ time complexity for a fixed number of iterations. We perform 1000 iterations for Fig. 4a and 4b.

# Data availability

The data used in this study are available from the corresponding authors upon request and can also be accessed by running the code provided as Supplementary Code S1.

# Code availability

The code used in this study is provided as Supplementary Code S1 to reproduce all data presented in this paper.

# References

1  Torquato, S. *Random heterogeneous materials: microstructure and macroscopic properties*. Vol. 16 (Springer, 2002).

2  Yu, S., Qiu, C.-W., Chong, Y., Torquato, S. & Park, N. Engineered disorder in photonics. *Nat. Rev. Mater.* **6**, 226-243 (2021).

3  Torquato, S. Hyperuniform states of matter. *Phys. Rep.* **745**, 1-95 (2018).

4  Batten, R. D., Stillinger, F. H. & Torquato, S. Classical disordered ground states: Super-ideal gases and stealth and equi-luminous materials. *J. Appl. Phys.* **104** (2008).

5  Man, W. *et al.* Photonic band gap in isotropic hyperuniform disordered solids with low dielectric contrast. *Opt. Express* **21**, 19972-19981 (2013).

6  Man, W. *et al.* Isotropic band gaps and freeform waveguides observed in hyperuniform disordered photonic solids. *Proc. Natl. Acad. Sci. U.S.A.* **110**, 15886-15891 (2013).

7  Uche, O. U., Stillinger, F. H. & Torquato, S. Constraints on collective density variables: Two dimensions. *Phys. Rev. E* **70**, 046122 (2004).

8  Lalau-Keraly, C. M., Bhargava, S., Miller, O. D. & Yablonovitch, E. Adjoint shape optimization applied to electromagnetic design. *Opt. Express* **21**, 21693-21701 (2013).

9  Molesky, S. *et al.* Inverse design in nanophotonics. *Nat. Photon.* **12**, 659-670 (2018).

10 Feichtner, T., Selig, O., Kiunke, M. & Hecht, B. Evolutionary optimization of optical antennas. *Phys. Rev. Lett.* **109**, 127701 (2012).

11 Huntington, M. D., Lauhon, L. J. & Odom, T. W. Subwavelength lattice optics by evolutionary design. *Nano Lett.* **14**, 7195-7200 (2014).

12 Yu, S. Evolving scattering networks for engineering disorder. *Nat. Comput. Sci.* **3**, 128-138 (2023).

13 Ma, W. *et al.* Deep learning for the design of photonic structures. *Nat. Photon.* **15**, 77-

90 (2021).

14	Yu, S., Piao, X. & Park, N. Machine learning identifies scale-free properties in disordered materials. *Nat. Commun.* **11**, 4842 (2020).

15	Bretto, A. *Hypergraph theory: An introduction*. Vol. 1 (Cham: Springer, 2013).

16	Battiston, F. *et al.* The physics of higher-order interactions in complex systems. *Nat. Phys.* **17**, 1093-1098 (2021).

17	Battiston, F. *et al.* Networks beyond pairwise interactions: Structure and dynamics. *Phys. Rep.* **874**, 1-92 (2020).

18	Rossi, M., Huber, M., Bruß, D. & Macchiavello, C. Quantum hypergraph states. *New J. Phys.* **15**, 113022 (2013).

19	Huang, J. *et al.* Demonstration of hypergraph-state quantum information processing. *Nat. Commun.* **15**, 2601 (2024).

20	Feng, Y., You, H., Zhang, Z., Ji, R. & Gao, Y. Hypergraph neural networks. *Proc. AAAI Conf. Artif. Intell.* **33**, 3558-3565 (2019).

21	Heydaribeni, N., Zhan, X., Zhang, R., Eliassi-Rad, T. & Koushanfar, F. Distributed constrained combinatorial optimization leveraging hypergraph neural networks. *Nat. Mach. Intell.* **6**, 664-672 (2024).

22	Ferraz de Arruda, G., Petri, G., Rodriguez, P. M. & Moreno, Y. Multistability, intermittency, and hybrid transitions in social contagion models on hypergraphs. *Nat. Commun.* **14**, 1375 (2023).

23	Sheng, A., Su, Q., Wang, L. & Plotkin, J. B. Strategy evolution on higher-order networks. *Nat. Comput. Sci.* **4**, 274-284 (2024).

24	Gonis, A. & Butler, W. H. *Multiple scattering in solids*. (Springer Science & Business Media, 1999).

25	Barabási, A.-L. *Network Science*.  (Cambridge University Press, 2016).

26	Barabási, A.-L. & Albert, R. Emergence of scaling in random networks. *Science* **286**, 509-512 (1999).

27	Schneidman, E., Berry, M. J., Segev, R. & Bialek, W. Weak pairwise correlations imply strongly correlated network states in a neural population. *Nature* **440**, 1007-1012 (2006).

28	Bairey, E., Kelsic, E. D. & Kishony, R. High-order species interactions shape ecosystem diversity. *Nat. Commun.* **7**, 12285 (2016).

29	Konstantinova, E. V. & Skorobogatov, V. A. Application of hypergraph theory in chemistry. *Discrete Math.* **235**, 365-383 (2001).

30	Morse, P. K., Kim, J., Steinhardt, P. J. & Torquato, S. Generating large disordered stealthy hyperuniform systems with ultrahigh accuracy to determine their physical properties. *Phys. Rev. Res.* **5**, 033190 (2023).

31	Shih, A., Casiulis, M. & Martiniani, S. Fast generation of spectrally shaped disorder. *Phys. Rev. E* **110**, 034122 (2024).

32	Saggio, V. *et al.* Experimental quantum speed-up in reinforcement learning agents. *Nature* **591**, 229-233 (2021).

33	Stanley, K. O. & Miikkulainen, R. Evolving neural networks through augmenting topologies. *Evol. Comput.* **10**, 99-127 (2002).

34	Stanley, K. O., Clune, J., Lehman, J. & Miikkulainen, R. Designing neural networks through neuroevolution. *Nat. Mach. Intell.* **1**, 24-35 (2019).

35	Orús, R. Tensor networks for complex quantum systems. *Nat. Rev. Phys.* **1**, 538-550 (2019).

## Acknowledgements

We acknowledge financial support from the National Research Foundation of Korea (NRF) through the Basic Research Laboratory (No. RS-2024-00397664), Innovation Research Center (No. RS-2024-00413957), Young Researcher Programs (No. 2021R1C1C1005031), and Midcareer Researcher Program (No. RS-2023-00274348), all funded by the Korean government. This work was supported by Creative-Pioneering Researchers Program and the BK21 FOUR program of the Education and Research Program for Future ICT Pioneers in 2024, through Seoul National University. We also acknowledge administrative support from SOFT foundry institute.

## Author contributions

K.P. and S.Yu conceived the idea. K.P. developed the theoretical tool and performed the numerical analysis. I.L., S.Youn, and G.L. examined the theoretical and numerical analysis. N.P. proposed the application of hybrid approach. S.Yu and N.P. supervised the findings of this work. All authors discussed the results and wrote the final manuscript.

## Competing interests

The authors have no conflicts of interest to declare.

## Additional information

**Correspondence and requests for materials** should be addressed to N.P. or S.Y.

# Figure Legends

**Fig. 1. Set operations for scattering hypergraphs. a**, Scattering from a multiparticle system. The black circle represents the multiparticle system. Green and cyan arrows represent incident and scattered waves with wavevector $\mathbf{k}_i$ and $\mathbf{k}_s$, respectively. The grey line indicates the structure factor of the system, while the lengths of the grey arrows describe the angular scattering intensity. **b**, The scattering network modelling of the system. Each dot and solid line represent a node and an edge, respectively. Only a small portion of the edges are shown. The transparency of the solid lines indicates the edge weights. The red dots represent the $n$th and $n'$th nodes, and the red solid line represents the edge connecting the two nodes. **c,d,h**, The seed set $\mathbf{Q}$ (**c**) decomposed into three subsets $\mathbf{Q}_1$, $\mathbf{Q}_2$, and $\mathbf{Q}_3$ (**d**) and the material phase $\mathbf{R}$ satisfying $\mathbf{R} = (\mathbf{Q}_1 + \mathbf{Q}_2) \cup \mathbf{Q}_3$ (**h**). **e-g**, Three types of set operations: the union (**e**), complement (**f**), and Minkowski addition (**g**).

**Fig. 2. Scattering hypergraph configuration. a**, A seed material $\mathbf{Q} = \mathbf{Q}_1 \cup \mathbf{Q}_2 \cup \mathbf{Q}_3$, allowing for generating various material phases through set operations, such as $\mathbf{R}$, $\mathbf{R'}$, and $\mathbf{R''}$. **b**, Wave scattering from $\mathbf{R} = (\mathbf{Q}_1 + \mathbf{Q}_2) \cup \mathbf{Q}_3$. Green and cyan arrows represent incident and scattered waves with wavevectors $\mathbf{k}_i$ and $\mathbf{k}_s$, respectively. **c**, The scattering hypergraph for $\mathbf{R}$. Because every group of nodes of type $i = (1,1), (2,2), (3,3), (1,2,3)$, or $(1,1,2,2)$ is linked by an $i$-type hyperedge, only a portion of hyperedges is visualized. For example, the (1,2,3)-type hyperedge connecting three nodes at $\mathbf{r}_n \in \mathbf{Q}_1$, $\mathbf{r}_{n'} \in \mathbf{Q}_2$, and $\mathbf{r}_{n''} \in \mathbf{Q}_3$ is illustrated. Its weight is defined as Eq. (4).

**Fig. 3. Hypergraph design of SHU materials. a,c**, Scattering hypergraphs before (**a**) and after (**c**) the attachment of a new node of $\mathbf{Q}_1$ from the 'Growth' process. The spatial configuration of the nodes represents the seed system $\mathbf{Q} = \mathbf{Q}_1 \cup \mathbf{Q}_2$. **b,d**, Generated materials $\mathbf{R} = \mathbf{Q}_1 + \mathbf{Q}_2$ before (**b**) and after (**d**) the 'Projection' process. **e,f**, Evolution of material structures of $\mathbf{Q}$ (**e**) and $\mathbf{R}$ (**f**) at $t = 20, 50$, and $198$. Because each subsystem evolves in alternating order, the particle number of $\mathbf{R}$ is $N = (t/2 + 1)^2$ for even epoch $t$. The boxes represent the supercell. **g,h**, Time evolutions of $\langle S(\mathbf{k}) \rangle_\mathbf{K}$ (**g**) and $S_i$ (**h**). 100 realizations are investigated with different Monte Carlo sampling of a real space (Methods).

**Fig. 4. Time complexity analysis. a,b**, Computation time for the design of an $N$-particle material (**a**) and resulting scattering response $\langle S(\mathbf{k}) \rangle_\mathbf{K}$ (**b**) using the collective-coordinate method (orange) and time evolution of scattering networks (green) and scattering hypergraphs (blue). **c**, Temporal variation of $\langle S(\mathbf{k}) \rangle_\mathbf{K}$ during the optimization process from the collective-coordinate method (orange) and the hybrid approach (purple). In the hybrid design approach, hypergraph evolution is terminated at the dotted vertical line and collective-coordinate optimization follows. The final particle number is set to $N = 500^2$ in both methods for comparison. Circular points and error bands denote the ensemble average and the standard deviation of an ensemble of 50 realizations for each design method in (**a-c**). See Methods for details.

**Supplementary Information for "Hypergraph modelling of wave scattering to speed-up material design"**

Kunwoo Park[1], Ikbeom Lee[1], Seungmok Youn[1], Gitae Lee[1], Namkyoo Park[2§], and Sunkyu Yu[1*]

[1]Intelligent Wave Systems Laboratory, Department of Electrical and Computer Engineering, Seoul National University, Seoul 08826, Korea

[2]Photonic Systems Laboratory, Department of Electrical and Computer Engineering, Seoul National University, Seoul 08826, Korea

E-mail address for correspondence: [§]nkpark@snu.ac.kr, [*]sunkyu.yu@snu.ac.kr

**Note S1. Derivation of scattering hypergraph models**

**Note S2. Correspondence between pairwise- and hyper-graph models**

**Note S3. Definition of the cost function**

**Note S4. Hypergraph evolutions for other set operations**

**Note S5. Preferential evolution of hyperedges**

**Note S6. Computational details in evaluating the cost function**

**Algorithm S1. Pseudo-code for hypergraph evolution**

**Note S1. Derivation of scattering hypergraph models**

The first-order scattering intensity of a multiparticle system $\mathbf{R} = \{\mathbf{r}_n: 1 \leq n \leq N\}$ is proportional to its structure factor $S(\mathbf{k}) = |\Sigma_n \exp(-i\mathbf{k}\cdot\mathbf{r}_n)|^2/N$. From Eq. (1) and (2) in the main text, the averaged scattering across $\mathbf{K}$ is proportional to:

$$\langle S(\mathbf{k}) \rangle_{\mathbf{K}} = 1 + \frac{2}{N} \sum_{n<n'} \langle \cos[\mathbf{k}\cdot(\mathbf{r}_n - \mathbf{r}_{n'})] \rangle_{\mathbf{K}}. \tag{S1}$$

This expression underpins the definition of edge weights in the previous study on pairwise-edge network modelling of wave scattering[1].

To generalize the above pairwise graph modelling to hypergraphs, we introduce a seed set $\mathbf{Q}$ and its mutually exclusive subsets $\mathbf{Q}_l = \{\mathbf{r}_n^l: 1 \leq n \leq N_l\}$ as defined in the main text. These subsets collectively determine the target material system $\mathbf{R}$ through set operations. Inspired by Eq. (S1), we express the scattering by $\mathbf{R}$ in the form of Eq. (3) in the main text, allowing for introducing a set of hyperedges that are determined by the applied set operations in constructing $\mathbf{R}$. In the following, we derive hyperedge weights and group weights, which characterize hyperedges, for three different examples of $\mathbf{R}$: (i) $\mathbf{R} = \mathbf{Q}_1 + \mathbf{Q}_2$, (ii) $\mathbf{R} = (\mathbf{Q}_1 + \mathbf{Q}_2) \cup \mathbf{Q}_3$, and (iii) $\mathbf{R} = (\mathbf{Q}_1 + \mathbf{Q}_2) \setminus \mathbf{Q}_3$. The resulting hypergraph for each example provides an alternative interpretation of wave scattering from $\mathbf{R}$ through the connectivity between hypergraph nodes that are the elements of $\mathbf{Q}$.

(i) $\mathbf{R} = \mathbf{Q}_1 + \mathbf{Q}_2$: By definition, the structure factor of $\mathbf{R}$ is

$$S(\mathbf{k}) = 1 + \frac{1}{N} \left\{ \sum_{n_2=1}^{N_2} \sum_{n_1 \neq n_1'} \cos\left[\mathbf{k}\cdot\left((\mathbf{r}_{n_1}^1 + \mathbf{r}_{n_2}^2) - (\mathbf{r}_{n_1'}^1 + \mathbf{r}_{n_2}^2)\right)\right] + \sum_{n_1=1}^{N_1} \sum_{n_2 \neq n_2'} \cos\left[\mathbf{k}\cdot\left((\mathbf{r}_{n_1}^1 + \mathbf{r}_{n_2}^2) - (\mathbf{r}_{n_1}^1 + \mathbf{r}_{n_2'}^2)\right)\right] \right.$$
$$\left. + \sum_{n_1 \neq n_1'} \sum_{n_2 \neq n_2'} \cos\left[\mathbf{k}\cdot\left((\mathbf{r}_{n_1}^1 + \mathbf{r}_{n_2}^2) - (\mathbf{r}_{n_1'}^1 + \mathbf{r}_{n_2'}^2)\right)\right] \right\}.$$

$$\tag{S2}$$

which results in the average across $\mathbf{K}$:

$$\langle S(\mathbf{k})\rangle_{\mathbf{K}} = 1 + \frac{2}{N}\left\{ N_2 \sum_{n_1 < n_1'} \left\langle \cos\left[\mathbf{k}\cdot\left(\mathbf{r}_{n_1}^1 - \mathbf{r}_{n_1'}^1\right)\right]\right\rangle_{\mathbf{K}} + N_1 \sum_{n_2 < n_2'} \left\langle \cos\left[\mathbf{k}\cdot\left(\mathbf{r}_{n_2}^2 - \mathbf{r}_{n_2'}^2\right)\right]\right\rangle_{\mathbf{K}} \right.$$
$$\left. + 2 \sum_{n_1 < n_1'} \sum_{n_2 < n_2'} \left\langle \cos\left[\mathbf{k}\cdot\left(\mathbf{r}_{n_1}^1 - \mathbf{r}_{n_1'}^1\right)\right]\cos\left[\mathbf{k}\cdot\left(\mathbf{r}_{n_2}^2 - \mathbf{r}_{n_2'}^2\right)\right]\right\rangle_{\mathbf{K}}\right\}. \quad (S3)$$

Comparing Eq. (S3) with Eq. (3) in the main text, hyperedge weights, group weights, and the number of hyperedges for each group are obtained as Table S1.

**Table S1. Hypergraph definitions of R = $Q_1$ + $Q_2$.** $i$ and $m$ denote the type and index of hyperedges, respectively, $M_i$ is the number of the $i$-type hyperedges, $w_m^i$ is the weight of the $m$th $i$-type hyperedge, and $\eta_i$ is the group weight of $i$-type hyperedges.

| R | $i$ | $w_m^i$ | $\eta_i$ | $M_i$ |
|---|---|---|---|---|
| | (1,1) | $\left\langle \cos\left[\mathbf{k}\cdot\left(\mathbf{r}_{n_1}^1 - \mathbf{r}_{n_1'}^1\right)\right]\right\rangle_{\mathbf{K}}$ | $N_2$ | $N_1(N_1-1)/2$ |
| $Q_1 + Q_2$ | (2,2) | $\left\langle \cos\left[\mathbf{k}\cdot\left(\mathbf{r}_{n_2}^2 - \mathbf{r}_{n_2'}^2\right)\right]\right\rangle_{\mathbf{K}}$ | $N_1$ | $N_2(N_2-1)/2$ |
| | (1,1,2,2) | $\left\langle \cos\left[\mathbf{k}\cdot\left(\mathbf{r}_{n_1}^1 - \mathbf{r}_{n_1'}^1\right)\right]\cos\left[\mathbf{k}\cdot\left(\mathbf{r}_{n_2}^2 - \mathbf{r}_{n_2'}^2\right)\right]\right\rangle_{\mathbf{K}}$ | 2 | $N_1(N_1-1)N_2(N_2-1)/4$ |

(ii) **R = ($Q_1$ + $Q_2$) ∪ $Q_3$**: The structure factor and its average across **K** become

$$S(\mathbf{k}) =$$
$$1 + \frac{1}{N}\left\{ \sum_{n_2=1}^{N_2} \sum_{n_1 \neq n_1'} \cos\left[\mathbf{k}\cdot\left(\left(\mathbf{r}_{n_1}^1 + \mathbf{r}_{n_2}^2\right) - \left(\mathbf{r}_{n_1'}^1 + \mathbf{r}_{n_2}^2\right)\right)\right] + \sum_{n_1=1}^{N_1} \sum_{n_2 \neq n_2'} \cos\left[\mathbf{k}\cdot\left(\left(\mathbf{r}_{n_1}^1 + \mathbf{r}_{n_2}^2\right) - \left(\mathbf{r}_{n_1}^1 + \mathbf{r}_{n_2'}^2\right)\right)\right] \right.$$
$$+ \sum_{n_3 \neq n_3'} \cos\left[\mathbf{k}\cdot\left(\mathbf{r}_{n_3}^3 - \mathbf{r}_{n_3'}^3\right)\right] + 2\sum_{n_1=1}^{N_1}\sum_{n_2=1}^{N_2}\sum_{n_3=1}^{N_3} \cos\left[\mathbf{k}\cdot\left(\left(\mathbf{r}_{n_1}^1 + \mathbf{r}_{n_2}^2\right) - \mathbf{r}_{n_3}^3\right)\right] \quad (S4)$$
$$\left. + \sum_{n_1 \neq n_1'} \sum_{n_2 \neq n_2'} \cos\left[\mathbf{k}\cdot\left(\left(\mathbf{r}_{n_1}^1 + \mathbf{r}_{n_2}^2\right) - \left(\mathbf{r}_{n_1'}^1 + \mathbf{r}_{n_2'}^2\right)\right)\right]\right\},$$

$$\langle S(\mathbf{k})\rangle_{\mathbf{K}} = 1 + \frac{2}{N}\left\{ N_2 \sum_{n_1<n_1'}\left\langle\cos\left[\mathbf{k}\cdot(\mathbf{r}_{n_1}^1 - \mathbf{r}_{n_1'}^1)\right]\right\rangle_{\mathbf{K}} + N_1 \sum_{n_2<n_2'}\left\langle\cos\left[\mathbf{k}\cdot(\mathbf{r}_{n_2}^2 - \mathbf{r}_{n_2'}^2)\right]\right\rangle_{\mathbf{K}}\right.$$

$$+ \sum_{n_3<n_3'}\left\langle\cos\left[\mathbf{k}\cdot(\mathbf{r}_{n_3}^3 - \mathbf{r}_{n_3'}^3)\right]\right\rangle_{\mathbf{K}} + \sum_{n_1=1}^{N_1}\sum_{n_2=1}^{N_2}\sum_{n_3=1}^{N_3}\left\langle\cos\left[\mathbf{k}\cdot(\mathbf{r}_{n_1}^1 + \mathbf{r}_{n_2}^2 - \mathbf{r}_{n_3}^3)\right]\right\rangle_{\mathbf{K}} \quad \text{(S5)}$$

$$\left. + 2\sum_{n_1<n_1'}\sum_{n_2<n_2'}\left\langle\cos\left[\mathbf{k}\cdot(\mathbf{r}_{n_1}^1 - \mathbf{r}_{n_1'}^1)\right]\cos\left[\mathbf{k}\cdot(\mathbf{r}_{n_2}^2 - \mathbf{r}_{n_2'}^2)\right]\right\rangle_{\mathbf{K}}\right\}.$$

Again comparing Eq. (S5) with Eq. (3) in the main text, we obtain hyperedge weights, group weights, and the number of hyperedges for each group in Table S2.

**Table S2. Hypergraph definitions of R = (Q$_1$ + Q$_2$) ∪ Q$_3$.**

| R | $i$ | $w_m^i$ | $\eta_i$ | $M_i$ |
|---|---|---|---|---|
| (Q$_1$ + Q$_2$) ∪ Q$_3$ | (1,1) | $\left\langle\cos\left[\mathbf{k}\cdot(\mathbf{r}_{n_1}^1 - \mathbf{r}_{n_1'}^1)\right]\right\rangle_{\mathbf{K}}$ | $N_2$ | $N_1(N_1-1)/2$ |
| | (2,2) | $\left\langle\cos\left[\mathbf{k}\cdot(\mathbf{r}_{n_2}^2 - \mathbf{r}_{n_2'}^2)\right]\right\rangle_{\mathbf{K}}$ | $N_1$ | $N_2(N_2-1)/2$ |
| | (3,3) | $\left\langle\cos\left[\mathbf{k}\cdot(\mathbf{r}_{n_3}^3 - \mathbf{r}_{n_3'}^3)\right]\right\rangle_{\mathbf{K}}$ | 1 | $N_3(N_3-1)/2$ |
| | (1,2,3) | $\left\langle\cos\left[\mathbf{k}\cdot(\mathbf{r}_{n_1}^1 + \mathbf{r}_{n_2}^2 - \mathbf{r}_{n_3}^3)\right]\right\rangle_{\mathbf{K}}$ | 1 | $N_1 N_2 N_3$ |
| | (1,1,2,2) | $\left\langle\cos\left[\mathbf{k}\cdot(\mathbf{r}_{n_1}^1 - \mathbf{r}_{n_1'}^1)\right]\cos\left[\mathbf{k}\cdot(\mathbf{r}_{n_2}^2 - \mathbf{r}_{n_2'}^2)\right]\right\rangle_{\mathbf{K}}$ | 2 | $N_1(N_1-1)N_2(N_2-1)/4$ |

(iii) **R = (Q$_1$ + Q$_2$) \ Q$_3$**: Similarly, we obtain

$$S(\mathbf{k}) =$$
$$1 + \frac{1}{N}\left\{2\sum_{n_3=1}^{N_3}\cos\left[\mathbf{k}\cdot(\mathbf{r}_{n_3}^3 - \mathbf{r}_{n_3'}^3)\right] + \sum_{n_2=1}^{N_2}\sum_{n_1\neq n_1'}\cos\left[\mathbf{k}\cdot\left((\mathbf{r}_{n_1}^1 + \mathbf{r}_{n_2}^2) - (\mathbf{r}_{n_1'}^1 + \mathbf{r}_{n_2}^2)\right)\right]\right.$$

$$+ \sum_{n_1=1}^{N_1}\sum_{n_2\neq n_2'}\cos\left[\mathbf{k}\cdot\left((\mathbf{r}_{n_1}^1 + \mathbf{r}_{n_2}^2) - (\mathbf{r}_{n_1}^1 + \mathbf{r}_{n_2'}^2)\right)\right] + \sum_{n_3\neq n_3'}\cos\left[\mathbf{k}\cdot(\mathbf{r}_{n_3}^3 - \mathbf{r}_{n_3'}^3)\right] \quad \text{(S6)}$$

$$\left. - 2\sum_{n_1=1}^{N_1}\sum_{n_2=1}^{N_2}\sum_{n_3=1}^{N_3}\cos\left[\mathbf{k}\cdot\left((\mathbf{r}_{n_1}^1 + \mathbf{r}_{n_2}^2) - \mathbf{r}_{n_3}^3\right)\right] + \sum_{n_1\neq n_1'}\sum_{n_2\neq n_2'}\cos\left[\mathbf{k}\cdot\left((\mathbf{r}_{n_1}^1 + \mathbf{r}_{n_2}^2) - (\mathbf{r}_{n_1'}^1 + \mathbf{r}_{n_2'}^2)\right)\right]\right\},$$

$$\langle S(\mathbf{k})\rangle_{\mathbf{K}} = 1 + \frac{2}{N}\left\{\sum_{n_3=1}^{N_3} 1 + N_2 \sum_{n_1<n_1'}\langle\cos[\mathbf{k}\cdot(\mathbf{r}_{n_1}^1 - \mathbf{r}_{n_1'}^1)]\rangle_{\mathbf{K}} + N_1 \sum_{n_2<n_2'}\langle\cos[\mathbf{k}\cdot(\mathbf{r}_{n_2}^2 - \mathbf{r}_{n_2'}^2)]\rangle_{\mathbf{K}}\right.$$

$$+ \sum_{n_3<n_3'}\langle\cos[\mathbf{k}\cdot(\mathbf{r}_{n_3}^3 - \mathbf{r}_{n_3'}^3)]\rangle_{\mathbf{K}} - \sum_{n_1=1}^{N_1}\sum_{n_2=1}^{N_2}\sum_{n_3=1}^{N_3}\langle\cos[\mathbf{k}\cdot(\mathbf{r}_{n_1}^1 + \mathbf{r}_{n_2}^2 - \mathbf{r}_{n_3}^3)]\rangle_{\mathbf{K}} \quad (S7)$$

$$\left.+ 2\sum_{n_1<n_1'}\sum_{n_2<n_2'}\langle\cos[\mathbf{k}\cdot(\mathbf{r}_{n_1}^1 - \mathbf{r}_{n_1'}^1)]\cos[\mathbf{k}\cdot(\mathbf{r}_{n_2}^2 - \mathbf{r}_{n_2'}^2)]\rangle_{\mathbf{K}}\right\}.$$

which results in the corresponding hypergraph parameters in Table S3.

**Table S3. Hypergraph definitions of $R = (Q_1 + Q_2) \setminus Q_3$.**

| R | $i$ | $w_m^i$ | $\eta_i$ | $M_i$ |
|---|---|---|---|---|
| (Q_1+Q_2) \ Q_3 | (3) | 1 | 1 | $N_3$ |
| | (1,1) | $\langle\cos[\mathbf{k}\cdot(\mathbf{r}_{n_1}^1 - \mathbf{r}_{n_1'}^1)]\rangle_{\mathbf{K}}$ | $N_2$ | $N_1(N_1-1)/2$ |
| | (2,2) | $\langle\cos[\mathbf{k}\cdot(\mathbf{r}_{n_2}^2 - \mathbf{r}_{n_2'}^2)]\rangle_{\mathbf{K}}$ | $N_1$ | $N_2(N_2-1)/2$ |
| | (3,3) | $\langle\cos[\mathbf{k}\cdot(\mathbf{r}_{n_3}^3 - \mathbf{r}_{n_3'}^3)]\rangle_{\mathbf{K}}$ | 1 | $N_3(N_3-1)/2$ |
| | (1,2,3) | $\langle\cos[\mathbf{k}\cdot(\mathbf{r}_{n_1}^1 + \mathbf{r}_{n_2}^2 - \mathbf{r}_{n_3}^3)]\rangle_{\mathbf{K}}$ | $-1$ | $N_1 N_2 N_3$ |
| | (1,1,2,2) | $\langle\cos[\mathbf{k}\cdot(\mathbf{r}_{n_1}^1 - \mathbf{r}_{n_1'}^1)]\cos[\mathbf{k}\cdot(\mathbf{r}_{n_2}^2 - \mathbf{r}_{n_2'}^2)]\rangle_{\mathbf{K}}$ | 2 | $N_1(N_1-1)N_2(N_2-1)/4$ |

**Note S2. Correspondence between pairwise- and hyper-graph models**

We establish the correspondence between the pairwise graph and hypergraph models of wave scattering. Consider an illustrative example of three seed subsets $\mathbf{Q}_l = \{\mathbf{r}_n^l: 1 \leq n \leq N_l\}$ with particle numbers $N_1 = 3$, $N_2 = 2$, and $N_3 = 2$, with the set operations $\mathbf{R} = (\mathbf{Q}_1 + \mathbf{Q}_2) \cup \mathbf{Q}_3$. Figure S1 illustrates the pairwise- and hyper-graph models for the material characterized by $\mathbf{R}$. To interpret the emergence of five hyperedge types and their weights and group weights, we show corresponding pairwise edges for each type of hyperedges. As shown, a hyperedge of type $i$ in the hypergraph model corresponds to $\eta_i$ edges, which is the group weight of $i$-type hyperedges, in the pairwise graph model, while the negative value of $\eta_i$ leads to the number of the removals of the pairwise edges. We note that such a one to multiple correspondence between hyperedges and pairwise edges is the origin of the superior compactness of the proposed hypergraph model. One can also easily verify that the weight of a hyperedge represents the average weight of the corresponding edges.

To obtain the explicit relationship between the numbers of hyperedges $M_i$ and pairwise edges $M$, we introduce the unweighted model by assuming the coincidence of all hypergraph nodes, which leads to the condition of

$$\sum_i \eta_i M_i = \sum_i \eta_i \sum_{m=1}^{M_i} 1 = \lim_{\forall \mathbf{r}_j^l \to 0} \sum_i \eta_i \sum_{m=1}^{M_i} w_m^i = \lim_{\forall \mathbf{r}_j^l \to 0} \frac{N}{2} \left( \langle S(\mathbf{k}) \rangle_{\mathbf{K}} - 1 \right) = \frac{N(N-1)}{2} = M. \quad (S8)$$

leading to $\Sigma_i \eta_i M_i = M$. As shown, the compactness of the hypergraph model is determined by the set of $\eta_i$ that can be found in Tables S1 to S3 for each example.

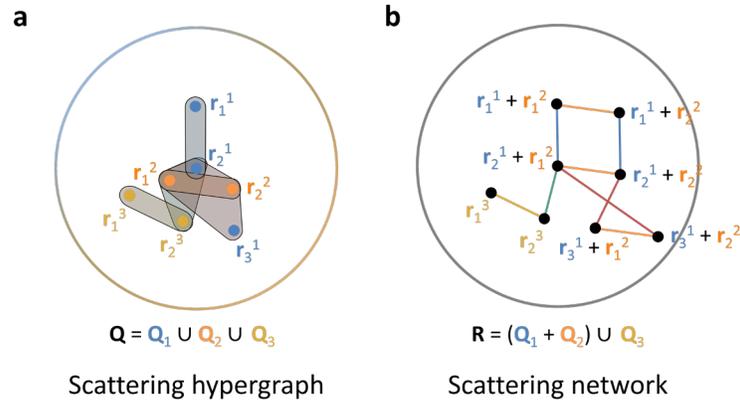

**Fig. S1. Correspondence between pairwise graph and hypergraph modelling of wave scattering. a,b**, The scattering hypergraph (**a**) and scattering pairwise network (**b**), which model wave scattering from $\mathbf{R} = (\mathbf{Q}_1 + \mathbf{Q}_2) \cup \mathbf{Q}_3$. Edge colour in (**b**) represents the interaction type of the corresponding hyperedge in (**a**).

**Note S3. Definition of the cost function**

Consider an arbitrary scattering hypergraph with hyperedges of weight $w_m^i$ ($1 \leq m \leq M_i$) and group weight $\eta_i$, which models wave response of the material **R** with $N$ particles by Eq. (3) in the main text. After the attachment of a node located at **r** in the $l$th subsystem, the material at the next epoch, denoted by **R'**, possesses different particle number $N'$ and new hyperedges emerging with hyperedge weights $w_m^i$ ($M_i < m \leq M_i'$). We note that the summation of weights for newly attached $i$-type hyperedges, by definition, equals the $i$-type hyperdegree of the attached node, $d_{\text{evol}}^i(\mathbf{r};l)$. Because the group weights can also depend on subsystem sizes (Tables S1-S3), we define the $i$-type group weight of the evolved hypergraph as $\eta_i'$. The scattering response of the material **R'** becomes:

$$\langle S_{\text{evol}}(\mathbf{k}) \rangle_\mathbf{K} = 1 + \frac{2}{N'} \sum_i \eta_i' \sum_{m=1}^{M_i'} w_m^i$$
$$= 1 + \frac{2}{N + \Delta N} \sum_i (\eta_i + \Delta \eta_i) \left( \sum_{m=1}^{M_i} w_m^i + d_{\text{evol}}^i(\mathbf{r};l) \right), \quad (S9)$$

where $\Delta N = N' - N$ and $\Delta \eta_i = \eta_i' - \eta_i$. Because $d_{\text{evol}}^i(\mathbf{r};l) = 0$ when $\Delta \eta_i \neq 0$ (Tables S1-S3),

$$\langle S_{\text{evol}}(\mathbf{k}) \rangle_\mathbf{K} = 1 + \frac{2}{N + \Delta N} \left\{ \sum_i (\eta_i + \Delta \eta_i) \sum_{m=1}^{M_i} w_m^i + \sum_i \eta_i d_{\text{evol}}^i(\mathbf{r};l) \right\}. \quad (S10)$$

By substituting Eq. (3) in the main text, we obtain

$$\langle S_{\text{evol}}(\mathbf{k}) \rangle_\mathbf{K} = 1 + \frac{2}{N + \Delta N} \left\{ \frac{N}{2} (\langle S(\mathbf{k}) \rangle_\mathbf{K} - 1) + \sum_i \Delta \eta_i \sum_{m=1}^{M_i} w_m^i + \rho(\mathbf{r};l) \right\}$$
$$= \frac{N \langle S(\mathbf{k}) \rangle_\mathbf{K} + \Delta N}{N + \Delta N} + \frac{2}{N + \Delta N} \left( \rho(\mathbf{r};l) + \sum_i \Delta \eta_i \sum_{m=1}^{M_i} w_m^i \right), \quad (S11)$$

for $\rho(\mathbf{r};l)$ defined in Eq. (5) in the main text. Equation (S11) shows that the minimization of $\rho(\mathbf{r};l)$ directly corresponds to the minimization of the structure factor and the corresponding scattering intensity.

**Note S4. Hypergraph evolutions for other set operations**

Figure S2 represents hypergraph evolutions for set operations $\mathbf{R} = (\mathbf{Q}_1 + \mathbf{Q}_2) \cup \mathbf{Q}_3$ and $\mathbf{R} = (\mathbf{Q}_1 + \mathbf{Q}_2) \setminus \mathbf{Q}_3$. Figures S2a-d illustrate the evolutions of $\mathbf{Q}$ and $\mathbf{R}$ for each set operation (See Videos S2 and S3 for the evolution with each set operation), showing similar behaviour of $\langle S(\mathbf{k}) \rangle_K$ to that of Fig. 3g in the main text due to the vanishing contribution of hyperedges of the types $i = (3), (3,3),$ and $(1,2,3)$ determined by $\mathbf{Q}_3$ for large $t$ (Fig. S2f and S2h). Because of the final subsystem sizes $N_1 = N_2 = 100$ and $N_3 = 99$ at $t = 297$, the scattering response of $\mathbf{R} = (\mathbf{Q}_1 + \mathbf{Q}_2) \cup \mathbf{Q}_3$ or $(\mathbf{Q}_1 + \mathbf{Q}_2) \setminus \mathbf{Q}_3$ is largely influenced by the $10^4$-particle system $\mathbf{Q}_1 + \mathbf{Q}_2$, rather than 99-particle system $\mathbf{Q}_3$.

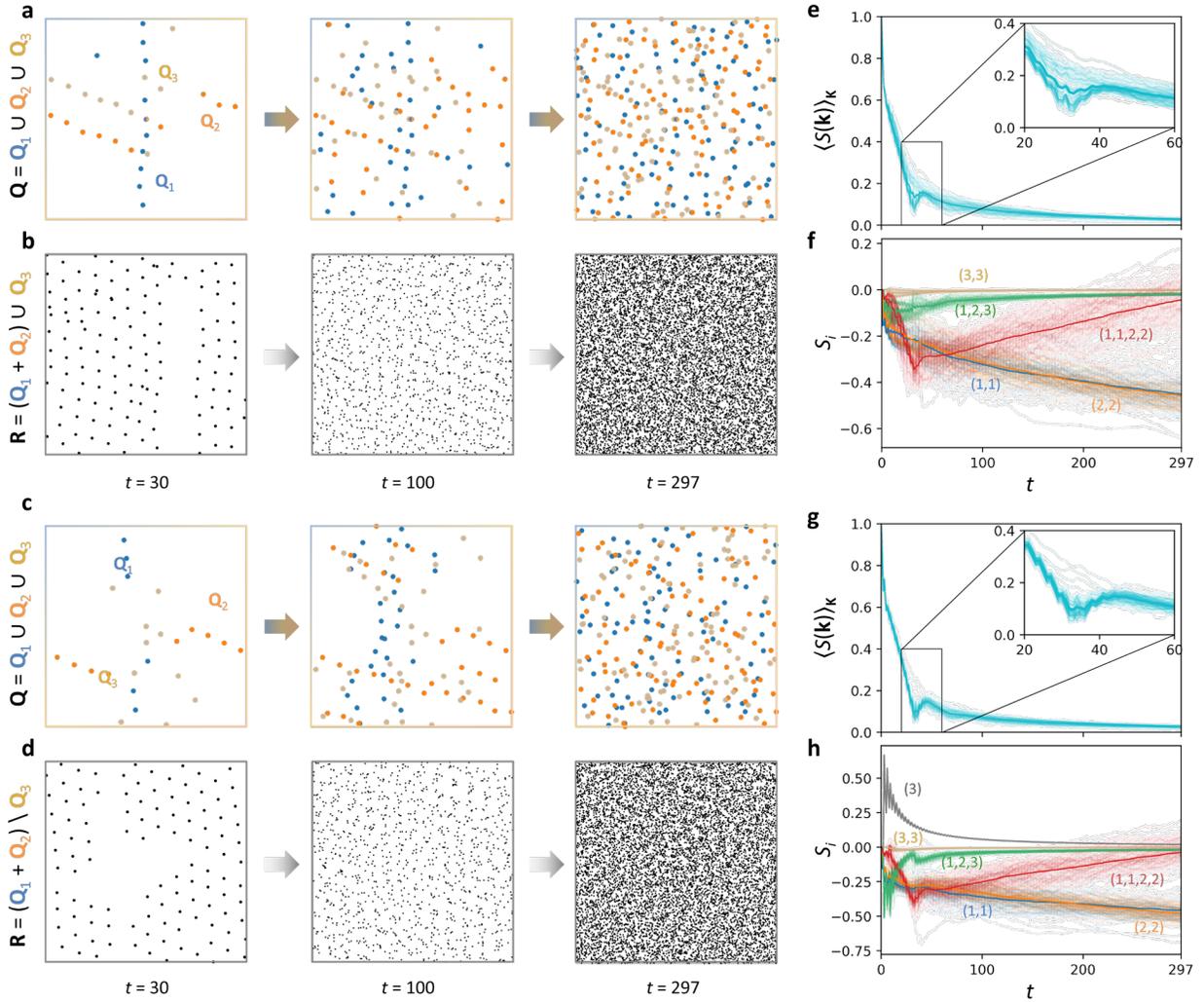

**Fig. S2. Hypergraph evolutions involving union and complement operations. a-d**, Evolutions of seed systems **Q** (**a,c**) and target materials **R** (**b,d**) for set operations **R** = (**Q**$_1$ + **Q**$_2$) ∪ **Q**$_3$ (**a,b**) and **R** = (**Q**$_1$ + **Q**$_2$) \ **Q**$_3$ (**c,d**) at the epoch $t$ = 30, 100, and 297. Each subsystem evolves in alternating order. The outer boxes represent the supercell boundary. **e-h**, Time evolution of $\langle S(\mathbf{k})\rangle_\mathbf{K}$ (**e,g**) and $S_i$ (**f,h**) for the set operations **R** = (**Q**$_1$ + **Q**$_2$) ∪ **Q**$_3$ (**e,f**) and **R** = (**Q**$_1$ + **Q**$_2$) \ **Q**$_3$ (**g,h**). 100 realizations are investigated with different initial sampling of real space (Methods).

**Note S5. Preferential evolution of hyperedges**

Figure S3 explores hypergraph phases of SHU materials by manipulating the hyperedge scattering, $S_{(1,1,2,2)}$, while targeting the suppression performance, $\langle S(\mathbf{k})\rangle_K$. We consider the modified cost function:

$$\rho(\mathbf{r};l) = \eta_{(1,1)}d_{evol}^{(1,1)}(\mathbf{r};l) + \eta_{(2,2)}d_{evol}^{(2,2)}(\mathbf{r};l) + \alpha \cdot \eta_{(1,1,2,2)}d_{evol}^{(1,1,2,2)}(\mathbf{r};l), \tag{S12}$$

where $\alpha$ denotes the preference of $S_{(1,1,2,2)}$ in the evolution: strengthen ($\alpha < 1$) or weaken ($\alpha > 1$) $S_{(1,1,2,2)}$ compared to the default value $\alpha = 1$. Figure S3a shows acceptable suppression performance ($\langle S(\mathbf{k})\rangle_K \lesssim 0.05$) for $\alpha$ within the range of [0.05, 1.05], where the contribution of the hyperedge $S_{(1,1,2,2)}$ to the entire scattering drastically differs from strong positive to strong negative (Fig. S3b). Substantial differences are also identified from the resulting materials **Q** and **R** for $\alpha = 0.05$ and 1.05 (Fig. S3c-f).

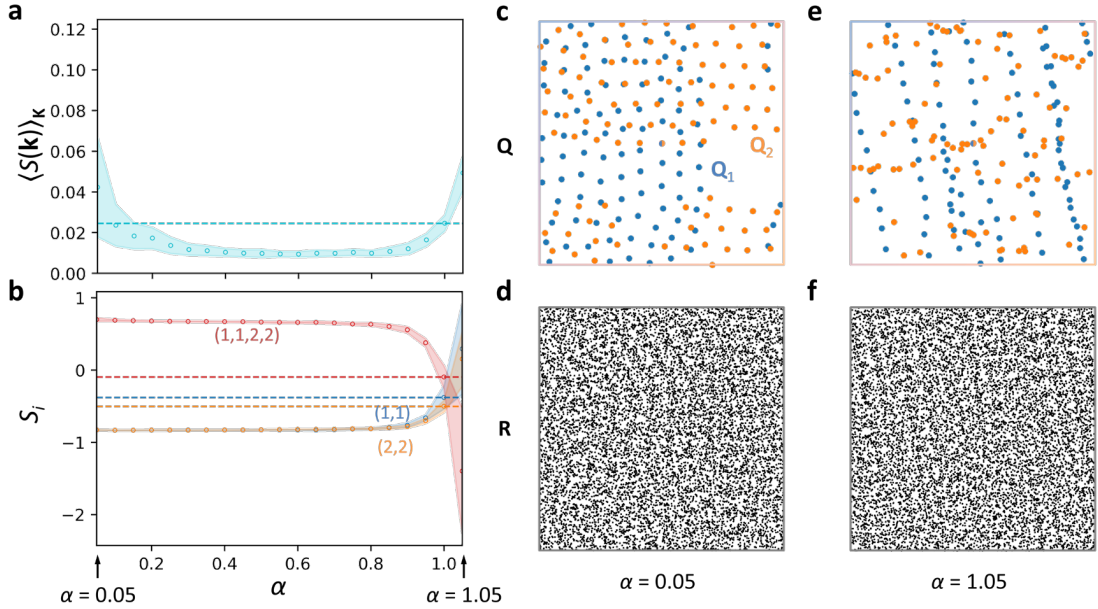

**Fig. S3. Preferential evolutions. a,b,** Scattering response $\langle S(\mathbf{k})\rangle_K$ (**a**) and $i$-type scattering $S_i$ (**b**) for different $\alpha$. Horizontal lines represent mean values from the default evolution process, $\alpha = 1$. 50 realizations are examined for each $\alpha$. **c-f,** A realization example of resulting seed systems **Q** (**c,e**) and the target material **R** (**d,f**) for $\alpha = 0.05$ (**c,d**) and $\alpha = 1.05$ (**e,f**).

**Note S6. Computational details in evaluating the cost function**

After $t$ epochs of hypergraph evolution for the set operation $\mathbf{R} = \mathbf{Q}_1 + \mathbf{Q}_2$, the particle number of the designed system becomes $N = O(t^2)$. Therefore, to maintain $O(N^{1/2})$ design complexity, computational cost for each epoch should remain constant regardless of the updated size of a hypergraph. Given the formula for the cost function in Eq. (5) of the main text, we need to compute the hyperdegree $d^i_{\text{evol}}(\mathbf{r};l)$ in $O(1)$. Although the number of hyperedges attached to a new node increases in time step $t$, we can achieve the required $O(1)$ efficiency by introducing auxiliary lookup maps $c_l(\mathbf{k})$ and $s_l(\mathbf{k})$ ($1 \leq l \leq L$, $\mathbf{k} \in \mathbf{K}$):

$$c_l(\mathbf{k}) = \sum_{\mathbf{r} \in \mathbf{Q}_l} \cos[\mathbf{k} \cdot \mathbf{r}], \tag{S13}$$

$$s_l(\mathbf{k}) = \sum_{\mathbf{r} \in \mathbf{Q}_l} \sin[\mathbf{k} \cdot \mathbf{r}], \tag{S14}$$

where $1 \leq l \leq L$ and $\mathbf{k} \in \mathbf{K}$. The time complexity for updating these lookup maps at each epoch, and the space complexity for their storage, only scale with the size of $\mathbf{K}$ rather than $\mathbf{Q}$ or $\mathbf{R}$. Using the lookup maps and basic trigonometric algebra, we derive following formulae for evaluating hyperdegrees of each order in $O(1)$ complexity.

(i) First-order ($i = (3)$)

$$d^{(3)}_{\text{evol}}(\mathbf{r};l) = \begin{cases} 0 & (l = 1, 2) \\ 1 & (l = 3) \end{cases} \tag{S15}$$

(ii) Second-order ($i = (l,l)$ for $l = 1, 2,$ or $3$)

$$d^{(l,l)}_{\text{evol}}(\mathbf{r};l') = \begin{cases} \langle \cos[\mathbf{k} \cdot \mathbf{r}] c_l(\mathbf{k}) + \sin[\mathbf{k} \cdot \mathbf{r}] s_l(\mathbf{k}) - 1 \rangle_{\mathbf{K}} & (l = l') \\ 0 & (l \neq l') \end{cases} \tag{S16}$$

(iii) Third-order ($i = (1,2,3)$)

$$d_{\text{evol}}^{(1,2,3)}(\mathbf{r};l) =$$
$$\begin{cases} \left\langle \cos[\mathbf{k}\cdot\mathbf{r}]\left(c_2(\mathbf{k})c_3(\mathbf{k})+s_2(\mathbf{k})s_3(\mathbf{k})\right)-\sin[\mathbf{k}\cdot\mathbf{r}]\left(s_2(\mathbf{k})c_3(\mathbf{k})-c_2(\mathbf{k})s_3(\mathbf{k})\right) \right\rangle_{\mathbf{K}} & (l=1) \\ \left\langle \cos[\mathbf{k}\cdot\mathbf{r}]\left(c_1(\mathbf{k})c_3(\mathbf{k})+s_1(\mathbf{k})s_3(\mathbf{k})\right)-\sin[\mathbf{k}\cdot\mathbf{r}]\left(s_1(\mathbf{k})c_3(\mathbf{k})-c_1(\mathbf{k})s_3(\mathbf{k})\right) \right\rangle_{\mathbf{K}} & (l=2) \\ \left\langle \cos[\mathbf{k}\cdot\mathbf{r}]\left(c_1(\mathbf{k})c_2(\mathbf{k})-s_1(\mathbf{k})s_2(\mathbf{k})\right)+\sin[\mathbf{k}\cdot\mathbf{r}]\left(s_1(\mathbf{k})c_2(\mathbf{k})-c_1(\mathbf{k})s_2(\mathbf{k})\right) \right\rangle_{\mathbf{K}} & (l=3) \end{cases}$$

(S17)

(iv) Fourth-order ($i = (1,1,2,2)$)

$$d_{\text{evol}}^{(1,1,2,2)}(\mathbf{r};l) = \begin{cases} \left\langle \left(\cos[\mathbf{k}\cdot\mathbf{r}]c_1(\mathbf{k})+\sin[\mathbf{k}\cdot\mathbf{r}]s_1(\mathbf{k})-1\right)\left(c_2^{\,2}(\mathbf{k})+s_2^{\,2}(\mathbf{k})-N_2\right)/2 \right\rangle_{\mathbf{K}} & (l=1) \\ \left\langle \left(\cos[\mathbf{k}\cdot\mathbf{r}]c_2(\mathbf{k})+\sin[\mathbf{k}\cdot\mathbf{r}]s_2(\mathbf{k})-1\right)\left(c_1^{\,2}(\mathbf{k})+s_1^{\,2}(\mathbf{k})-N_1\right)/2 \right\rangle_{\mathbf{K}} & (l=2) \\ 0 & (l=3) \end{cases}$$

(S18)

**Algorithm S1. Pseudo-code for hypergraph evolution.**

---

1:    Initialize the governing set operations

2:    Initialize the total evolution steps $t_{max}$

3:    Initialize the number of discretized points $n_{disc}$

4:    Initialize the material systems **Q** and **R**

5:    Initialize the scattering hypergraph parameters

6:    Initialize the reciprocal space of interest **K** for the evolution process

7:    Initialize the auxiliary lookup maps $c_l(\mathbf{k})$ and $s_l(\mathbf{k})$

8:    **for** every $t$ where $1 \leq t \leq t_{max}$ **do**

9:        Select the subsystem $l$ to evolve

10:       **if** ($\mathbf{R} = (\mathbf{Q}_1 + \mathbf{Q}_2) \setminus \mathbf{Q}_3$) and ($l = 3$) **do**

11:           Define candidate positions $\mathbf{\Omega}_{cand} = \mathbf{R}$

12:       **else do**

13:           Define candidate positions $\mathbf{\Omega}_{cand}$ through Monte Carlo discretization

14:       Calculate hyperdegrees $d^i{}_{evol}(\mathbf{r};l)$ for each candidate position **r** using Eq. (S15-18)

15:       Calculate cost functions $\rho(\mathbf{r};l)$ for each candidate position **r** using Eq. (5) or (S12)

16:       Select $\mathbf{r} = \mathrm{argmin}\, \rho(\mathbf{r};l)$ for $\mathbf{r} \in \mathbf{\Omega}_{cand}$

17:       Update the material systems **Q** and **R**

18:       Update the scattering hypergraph parameters using Eq. (3)

19:       Update the auxiliary lookup maps $c_l(\mathbf{k})$ and $s_l(\mathbf{k})$

20: **end for**

# References

1  Yu, S. Evolving scattering networks for engineering disorder. *Nat. Comput. Sci.* **3**, 128-138 (2023).